\documentclass[aps,prl,reprint,superscriptaddress]{revtex4-2}
\usepackage{graphicx}
\usepackage{xcolor}
\usepackage{dcolumn}
\usepackage{bm}
\usepackage[utf8]{inputenc}

\begin{document}
	
	\preprint{APS/123-QED}
	
	\title{Direct Evidence for Robust Bulk Band Gap Across the Charge Density Wave Transition in TiSe$_2$}
	
	\author{Turgut Yilmaz}
	\email{trgt2112@gmail.com}
	\affiliation{Department of Physics, Xiamen University Malaysia, Sepang 43900, Malaysia}
	\affiliation{Department of Physics, University of Connecticut, Storrs, CT 06269, USA}

	\author{Anil Rajapitamahuni}
	\affiliation{Department of Physics, SRM University - AP, Amaravati, Andhra Pradesh, 522502, India}
	\affiliation{Department of Physics and Astronomy, University of Nebraska-Lincoln, Nebraska 68588, USA}

	\author{Muhammad Awais Fiaz}
	\affiliation{Department of Physics and Astronomy, University of New Hampshire, Durham, NH 03824, USA}
	
	\author{Zhenxian Liu}
	\affiliation{National Synchrotron Light Source II, Brookhaven National Lab, Upton, New York 11973, USA}
	
	\author{Jerzy T. Sadowski}
	\affiliation{Center for Functional Nanomaterials, Brookhaven National Lab, Upton, New York, 11973, USA}
	
	\author{Abdullah Al-Mahboob}
	\affiliation{Center for Functional Nanomaterials, Brookhaven National Lab, Upton, New York, 11973, USA}

	\author{Asish K. Kundu}
	\affiliation{National Synchrotron Light Source II, Brookhaven National Lab, Upton, New York 11973, USA}

	\author{Shawna M. Hollen}
	\affiliation{Department of Physics and Astronomy, University of New Hampshire, Durham, NH 03824, USA}

	\author{Boris Sinkovic}
	\affiliation{Department of Physics, University of Connecticut, Storrs, CT 06269, USA}
	
	\author{Elio Vescovo}
	\affiliation{National Synchrotron Light Source II, Brookhaven National Lab, Upton, New York 11973, USA}

	\date{\today}

\begin{abstract}
	
The mechanism driving the charge density wave (CDW) transition in TiSe$_2$ has been debated for decades, with proposals ranging from an excitonic insulator to a lattice-driven instability. A central question remains whether the transition involves an opening or enhancement of the bulk band gap. Using high-resolution angle-resolved photoemission spectroscopy, we directly track the temperature evolution of the bulk band edges across the CDW transition at $T{\text{CDW}}\approx 200$ K. Contrary to the expectation of a gap-opening transition, we find that the size of the fundamental band gap remains constant from the high-temperature normal phase down to 160 K. While the CDW induces clear band-folding signatures and spectral weight redistribution, the underlying band extrema are unperturbed. These results demonstrate that TiSe$_2$ does not undergo a temperature-driven electronic gap opening. Instead, they support a scenario where the transition is governed by a lattice symmetry-breaking reconstruction that folds, but does not gap, the electronic structure of a pre-existing band insulator.

\end{abstract}

\maketitle


\section{Introduction}

Charge density waves (CDWs) represent one of the most ubiquitous collective phenomena in condensed matter, arising from the delicate interplay between electrons and the lattice~\cite{gruner2018density,johannes2008fermi,wilson1975charge}. While the Peierls mechanism provides an elegant framework for CDWs in one-dimensional metals, many real materials, including the extensively studied transition metal dichalcogenide TiSe$_2$ defy this simple picture. TiSe$_2$ presents a fundamental puzzle. It is a narrow gap semiconductor, not a metal, yet undergoes a CDW transition~\cite{rossnagel2011origin}. This has sparked decades of debate: is this an exotic 'excitonic insulator' driven purely by electronic correlations, or a conventional lattice-driven instability? Despite intensive study, a definitive test has remained elusive, whether the electronic gap itself opens or changes at the transition~\cite{hughes1977structural,kidd2002electron,rossnagel2011origin,knowles2020fermi}.

TiSe$_2$ undergoes a commensurate $2\times2\times2$ CDW transition at $T_{\mathrm{CDW}} \approx 200$~K, accompanied by a periodic lattice distortion (PLD) and Brillouin-zone reconstruction~\cite{di1976electronic}. Early interpretations attributed the transition to an excitonic insulator mechanism, in which Coulomb-driven electron, hole pairing induces a temperature-dependent hybridization gap~\cite{rossnagel2002charge,cercellier2007evidence,monney2011exciton,monney2012mapping,kogar2017signatures}. In contrast, alternative scenarios emphasize lattice-driven symmetry breaking and band folding, describing TiSe$_2$ as a small-gap semiconductor whose electronic structure is reconstructed but not fundamentally gapped by the transition~\cite{weber2011electron,calandra2011charge,hedayat2019excitonic}. Despite extensive theoretical and experimental efforts, whether a genuine electronic gap opens or evolves at $T_{\mathrm{CDW}}$ remains unsettled. While previous ARPES studies reported either apparent gap enhancement or gap reduction in different temperature regimes~\cite{chen2015charge,chen2016hidden,watson2019orbital}, a direct temperature-dependent determination of the bulk band extrema across both the $\Gamma$-$M$ and $A$-$L$ planes has remained unavailable. In particular, earlier studies generally inferred the gap indirectly by comparing spectral features measured at different momentum locations, photon energies, or experimental geometries. Here, by simultaneously resolving the folded valence and conduction band edges within the same spectra across multiple $k_z$ planes, we directly demonstrate that the bulk band gap remains robust across the primary CDW transition.

A decisive test of these competing pictures requires direct determination of the bulk band gap across the transition. If the CDW is driven by excitonic condensation, one expects a measurable enhancement or opening of a hybridization gap at $T_{\mathrm{CDW}}$. Conversely, a lattice-driven reconstruction should primarily induce band folding and spectral redistribution without significant modification of the underlying band extreme.

Here we present high-resolution ARPES measurements across the CDW transition in TiSe$_2$, employing multiple photon energies and polarization geometries to isolate the conduction and folded valence bands. We directly track the temperature evolution of the band edges from 300~K to 160~K. The transition manifests solely through the emergence of folded spectral weight, with no shift of the band extrema. These results rule out a temperature-driven excitonic gap at $T_{\mathrm{CDW}}$ and establish that the primary CDW order in TiSe$_2$ arises from lattice-driven symmetry breaking of a pre-existing narrow-gap semiconducting state.

\begin{figure*}[t]
	\centering
	\includegraphics[width=0.75\textwidth]{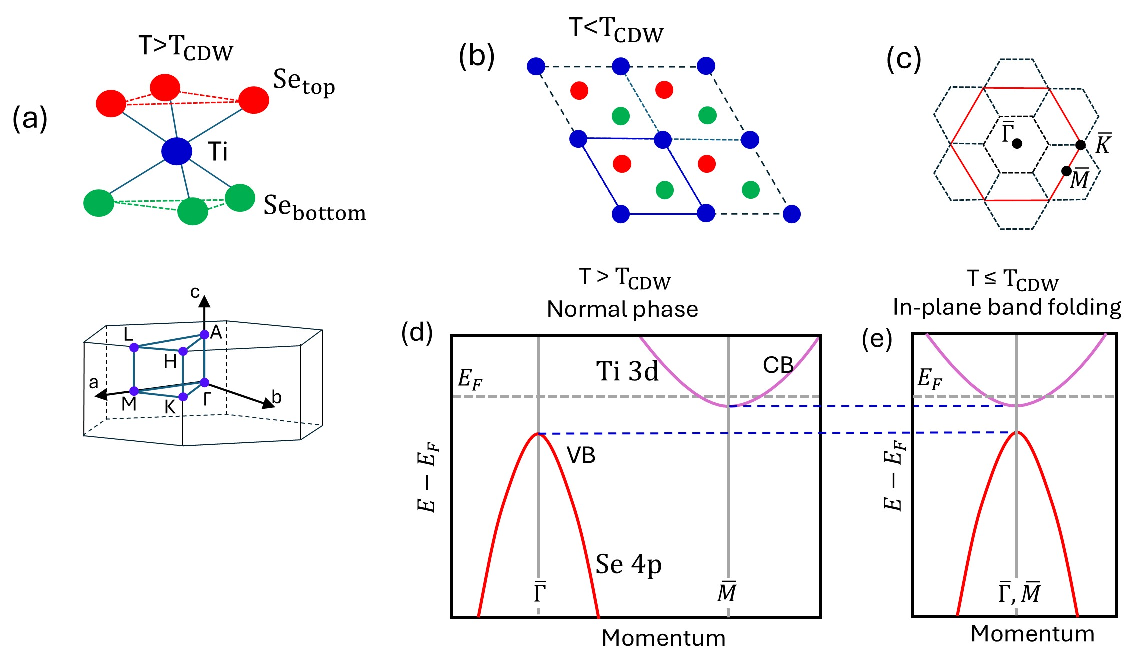} 
	
	\caption{\label{fig:epsart} 
		(a) Crystal structure of TiSe$_2$ in the normal phase (upper panel) and the three-dimensional Brillouin zone showing the high-symmetry points $\Gamma$, $A$, $M$, $L$, $K$, and $H$ (lower panel). (b) In-plane \(2a \times 2a\) periodic lattice distortion in the CDW phase. (c) Brillouin zone of the normal phase (red hexagon) and the CDW superstructure (dashed black hexagons). (d) Schematic electronic band structure of the normal phase. (e) Electronic structure in the CDW phase between 160~K and T$_{\text{CDW}}$. Blue lines connect panels (d) and (e) to indicate that the bulk band gap remains unchanged across T$_{\text{CDW}}$.
	}
	
\end{figure*}

High-resolution angle-resolved photoemission spectroscopy (ARPES) measurements were performed on high-quality single crystals of 1T-TiSe$_2$ (obtained from 2Dsemiconductors) at the ESM beamline (21-ID-1) of NSLS-II using a DA30 Scienta electron spectrometer~\cite{rajapitamahuni2024electron}. Samples were cleaved in situ at pressures below $3\times10^{-11}$~Torr, with a photon beam spot size of approximately $5~\mu\mathrm{m}^2$. Measurements employed multiple photon energies (119~eV and 95~eV) and polarizations (linear vertical and circular left) to selectively probe different orbital symmetry components. Sample temperature was monitored by a calibrated silicon diode with an absolute accuracy of $\pm$3~K. The band gap was extracted by fitting energy distribution curves at the valence and conduction band edges. The high crystalline quality of the samples was independently confirmed by low-temperature scanning tunneling microscopy and room-temperature low energy electron diffraction measurements presented in the Supplementary Information Figure~S1.

Figure~1 presents a schematic of the temperature dependent structural and electronic evolution of TiSe$_2$ across the CDW transition. Figure 1(a) depicts the atomic structure above T$_{\text{CDW}}$ (upper panel) together with the three-dimensional Brillouin zone (lower panel), while the CDW-distorted crystal structure below T$_{\text{CDW}}$ is shown in Figure 1(b). In the high temperature phase (T $>$ T$_{\text{CDW}}$), TiSe$_2$ retains its undistorted 1T structure, with Se$_\mathrm{top}$, Ti, and Se$_\mathrm{bottom}$ atoms arranged in a centrosymmetric trigonal configuration. Upon cooling below T$_{\text{CDW}}$, a 2$\times$2$\times$2 PLD emerges, leading to a symmetry breaking distortion of the lattice. This structural modulation reconstructs the Brillouin zone, as illustrated in Figure~1(c), where the original hexagonal zone is reduced due to the emergent superlattice periodicity.

Figures~1(d–e) illustrate the evolution of the electronic structure based on our ARPES data which will be discussed in the proceeding sections. In the normal phase (Figure~1(d)), the Se~4p-derived valence band peaks at the $\overline{\Gamma}$ point, while the Ti~3d conduction band has its minimum at the $\overline{M}$ point. Here, the normal phase electronic structure with a narrow gap is experimentally consistent with the earlier ARPES reports~\cite{huber2024ultrafast,watson2019orbital,yilmaz2023gapless,ou2024incoherence}. Representative normal-phase ARPES spectra measured along the $A$-$L$ and $\Gamma$-$M$ directions are presented in the Supplementary Information Figure~S2. As the temperature decreases below T$_{\text{CDW}}$, Figure~1(e)), band folding takes place, bringing $\overline{\Gamma}$-point valence states into the  $\overline{M}$ region. In this schematic representation, horizontal blue lines represent that the bulk band gap remains nearly constant during this process.

\begin{figure*}[t]
	\centering
	\includegraphics[width=0.9\textwidth]{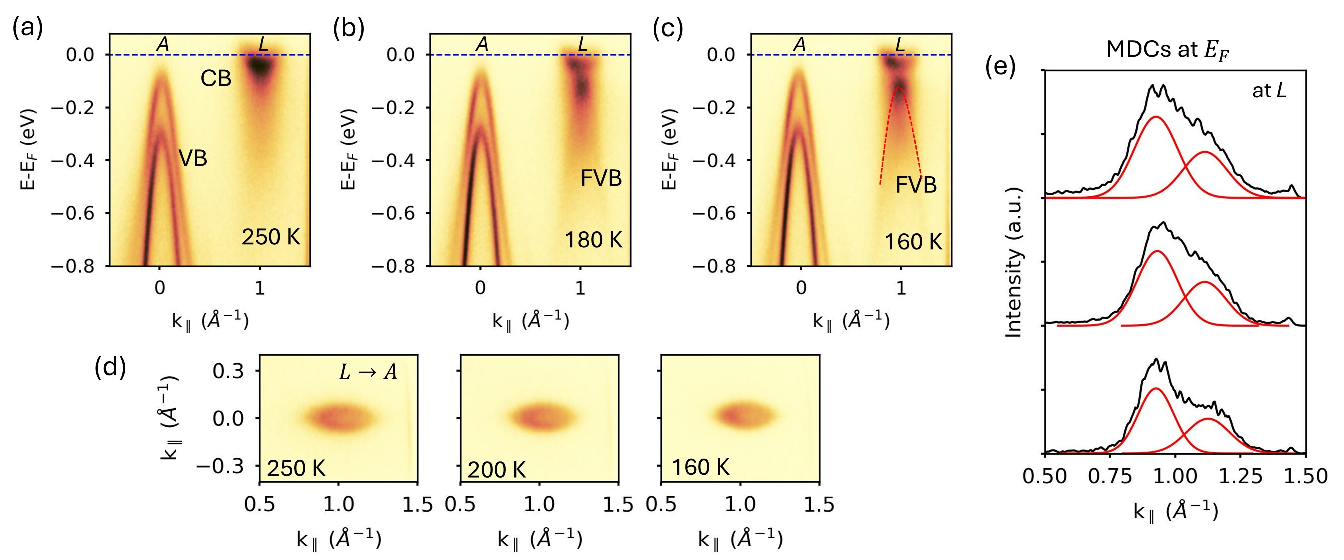} 
	
	\caption{\label{fig:epsart} 
		(a–c) ARPES intensity maps along the $A$-$L$ cut at various temperatures. Above T$_{CDW}$ (250~K), a dispersive valence band (VB) and a broad conduction band (CB) are observed. Between T$_{CDW}$ and ~160~K, a folded valence band (FVB) emerges at the $L$-point (marked with red dashed parabola) while the CB remains largely unchanged. (d) Corresponding Fermi-surface maps around the $L$-point. Elliptical contours persist, even across the T$_{CDW}$ and down to 160~K. (e) Momentum-distribution curves (MDCs) at the Fermi level ($E_F$) extracted at the L-point from panels (d). Red curves represent Gaussian fits to the conduction-band spectral feature. The momentum position remains nearly unchanged across temperature.
	}
	
\end{figure*}

Figure~2 presents the temperature evolution of the band structure measured with 119~eV photons along the $A$--$L$ plane. At 250~K, above $T_{\mathrm{CDW}}$, the spectra display a highly dispersive valence band near $A$ and a broad conduction band near $L$ (Figure~2(a)). Upon cooling to 180~K and 160~K [Figures~2(b,c)], a folded valence band (FVB) appears at the $M$ point, while the overall band dispersions, particularly the conduction band, remain nearly unchanged despite still being well below $T_{\mathrm{CDW}}$.

The transition is more clearly reflected in the temperature-dependent Fermi surface maps (Figure~2(d)). At 250 K, the normal phase Fermi surface near $L$ consists of an ellipsoidal electron pocket derived from Ti~3$d$ states. At 200~K and 160~K, these contours remain nearly identical, indicating that the initial CDW transition primarily introduces band folding and spectral weight redistribution without notably modifying the Fermi surface topology. Furthermore, Figure~2(e) shows momentum distribution curves (MDCs) at $E_F$ extracted near the L-point. The MDC peak positions remain nearly unchanged across temperature, further confirming the robustness of the conduction band spectral feature through the transition.

Having established the robustness of the Fermi surface across $T_{\mathrm{CDW}}$, we now directly examine the evolution of the bulk band gap. To probe the bulk band gap across the CDW transition temperature T$_{\text{CDW}}$, we present two complementary ARPES datasets: one recorded with 119~eV linear vertical (LV) polarization along the $L$-$L'$ direction, and another with 95~eV circular left (CR) polarized light along the $M$–$\Gamma$ direction in Figure~3(a, b), respectively. These configurations provide optimal spectral weight for resolving both the FVB and conduction band at $M$($L$)-point. In both datasets, the ARPES spectra between 160~K and T$_{\text{CDW}}$ appear nearly identical, characterized by a relatively broad FVB and conduction band at $M$($L$)-point.

Figures~4(a) show energy-distribution curves (EDCs) at the $L$-point, extracted from Figure~3(a) and compared with the room temperature reference spectrum. Upon cooling, the FVB first appears as a weak shoulder at 200~K and evolves into a distinct peak at 160~K, yet its binding energy remains aligned with the normal phase conduction band. This invariance demonstrates that neither the conduction nor valence band edge exhibits any measurable shift across $T_{\mathrm{CDW}}$. Figure~4(b) further supports this conclusion. The FVB gradually gains spectral weight, with a weak shoulder already visible at 220~K due to CDW fluctuations, and forms a clear peak near 110~meV at 160~K. Consistent with the $L$-point data, EDCs at the $M$-point likewise show no detectable movement of the band extrema across $T_{\mathrm{CDW}}$.

To quantify this invariance, we extracted the bulk gap independently at $k_z = A$ and $k_z = \Gamma$ (Figures~4(c,d)) In both cases, the gap magnitude remains in the range of 85–100~meV across the measured temperature window and does not exhibit an abrupt enhancement at $T_{\mathrm{CDW}}$. The absence of a systematic or momentum dependent gap opening demonstrates that the observed behavior is robust throughout the three dimensional Brillouin zone. The small variations between temperatures are comparable to the fitting uncertainty and do not follow a coherent trend across $k_z$, indicating that they do not reflect a thermodynamic order-parameter-like evolution. These results, which demonstrate the robustness of the bulk band gap across the CDW transition, provide direct experimental resolution to the long-standing debate in TiSe$_2$, clarifying that the primary order arises from lattice-driven band folding rather than a temperature-dependent electronic gap opening. The invariance of the bulk band gap across $T_{\mathrm{CDW}}$ highlights the primacy of structural modulation, consistent with theoretical analyses that identify momentum-dependent electron–phonon coupling as the stabilizing mechanism of charge order in prototypical CDW systems \cite{johannes2008fermi}. Ab initio studies further support this view, showing that the instability in TiSe$_2$ corresponds to a symmetry-lowering structural phase transition reinforced by orbital polarization \cite{zhu2012origin}.

Our finding of a nearly constant band gap across $T_{\mathrm{CDW}}$ stands in contrast to earlier ARPES studies~\cite{chen2016hidden,chen2015charge}, which reported notable gap widening across the transition. The discrepancy likely reflects methodological differences. In these works, gap values were obtained by separately tracking valence-band edges at $\Gamma$ relative to the conduction minimum at $L$. In our measurements, valence- and conduction-band edges are simultaneously resolved within single EDCs at the same $k$-point, using optimized photon energies and polarizations to maximize spectral weight, thereby eliminating uncertainties from cross-referencing. Moreover, recent high-resolution ARPES study has revealed additional electronic states at $\overline{\Gamma}$~\cite{yilmaz2025orbital} that were not accounted for in the earlier studies. In particular, one valence-band branch near $\Gamma$ exhibits strong orbital-selective temperature-dependent renormalization, whereas other valence states remain comparatively unaffected. Since this strongly renormalized branch is not clearly resolved at the $M(L)$ point, direct comparison of separate EDCs measured at $\Gamma(A)$ and $M(L)$ may not always provide a reliable determination of the intrinsic bulk-gap evolution. Subsequent ARPES and theoretical works~\cite{watson2019orbital,yilmaz2023gapless,pashov2025tise2} also consistently report a much smaller low-temperature gap deep in the CDW phase.

\begin{figure}[t]
	\centering
	\includegraphics[width=0.45\textwidth]{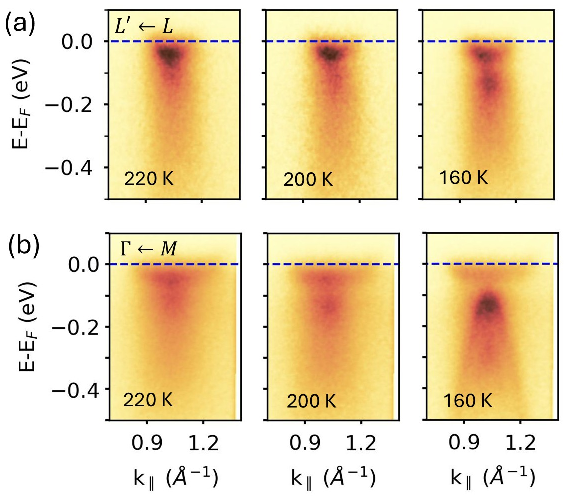} 
	
	\caption{\label{fig:epsart} 
		(a,b) ARPES intensity maps measured with 119 eV linear vertical (LV) and 95 eV circular left (CL) polarized light, sensitive to the folded valence band (FVB) and conduction band near the $L$-point and along the $\Gamma$-$M$, respectively. 
	}
	
\end{figure}

\begin{figure}[htp]
	\centering
	\includegraphics[width=0.45\textwidth]{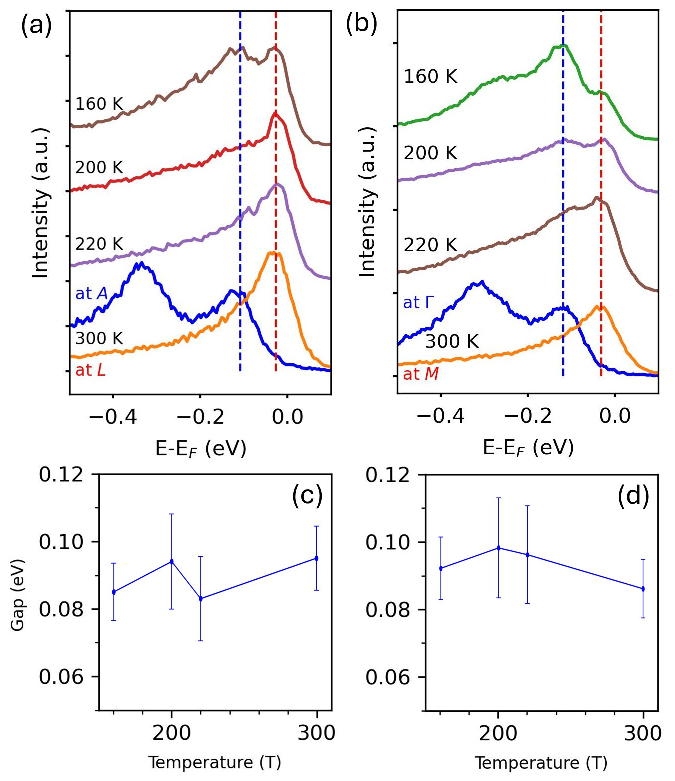} 
	
	\caption{\label{fig:epsart} 
		(a,b) Energy-distribution curves (EDCs) at the $L$- and $M$-points obtained from Figure 3(a,b), respectively. Blue and red dashed lines mark the valence and conduction band, respectively. EDCs at room temperature are obtained with 119 eV and 95 eV linear vertical (LV) polarized lights. (c,d) Gap versus temperature plots obtained from (a,b), respectively. Peak positions are extracted from fits to the spectra, given in Supplementary Information Figure~S3. Error bars represent the standard deviation of the fitted peak positions obtained from the EDC analysis.
	}
	
\end{figure}

Our finding of a nearly constant band gap across $T_{\mathrm{CDW}}$ stands in contrast to an earlier ARPES studies~\cite{chen2016hidden,chen2015charge}, which reported notable gap widening across the transition. The discrepancy likely reflects methodological differences. In these works, gap values were obtained by separately tracking valence band edges at $\Gamma$ relative to the conduction minimum at $L$. In our measurements, valence and conduction band edges are simultaneously resolved within single EDCs at the same $k$-point, using optimized photon energies and polarizations to maximize spectral weight, thereby eliminating uncertainties from cross referencing. Moreover, recent high-resolution laser ARPES has revealed additional electronic states at $\overline{\Gamma}$~\cite{yilmaz2025orbital} that were not accounted for in the earlier studies, while subsequent ARPES and theoretical works~\cite{watson2019orbital,yilmaz2023gapless,pashov2025tise2} consistently report a much smaller low-temperature gap deep in the CDW phase.

In conclusion, our results provide direct experimental evidence resolving the long standing debate over the nature of the CDW transition in TiSe$_2$. By tracking the bulk band edges across $T_{\mathrm{CDW}}$, we show that the transition does not involve an electronic gap modification, but instead reflects a lattice driven reconstruction of a pre-existing narrow-gap semiconductor. This interpretation is supported by recent theoretical work identifying TiSe$_2$ as a symmetry-breaking band insulator, in which excitonic effects are not the primary driver of the transition~\cite{pashov2025tise2}.

The observed gap invariance cannot be attributed to extrinsic effects such as chemical-potential shifts, sample quality, or measurement conditions. In the distorted phase, the folded valence band and conduction band are simultaneously resolved within single EDCs, allowing the bulk gap to be determined directly. Any rigid chemical potential shift would translate all spectral features uniformly and therefore leave the extracted gap unchanged. Likewise, variations in stoichiometry or disorder would affect spectra similarly above and below $T_{\mathrm{CDW}}$ within our narrow temperature window and cannot account for the constant gap we observe. Furthermore, the high crystalline quality of our samples, combined with micron-scale beam spots and high energy–momentum resolution, ensures that the extracted band edges reflect the intrinsic bulk electronic structure of TiSe$_2$.

While the onset of CDW order at $T_{\mathrm{CDW}}$ induces the band folding discussed above, TiSe$_2$ exhibits additional electronic reconstruction at lower temperatures of $\approx$ 160 K~\cite{ou2024incoherence,yilmaz2025orbital}. Previous ARPES studies near $\sim$20~K report a qualitative transformation of the conduction band from a broad, weakly dispersive feature into a sharp, V-shaped dispersion, accompanied by a reduced bulk gap~\cite{watson2019orbital}. Such pronounced renormalization is unlikely to arise from a smooth evolution of the primary CDW order parameter and instead points to a distinct low-temperature reconstruction. Additional structural and spectroscopic studies likewise indicate secondary transitions near $\sim$160~K, including changes in Fermi-surface topology and spectral coherence~\cite{amin2024two,diego2024electronic,guo2025plane}. Consistent with this picture, temperature-dependent Raman measurements presented in the Supplementary Information Figure S4 show that the CDW related phonon response remains relatively weak near $T_{\mathrm{CDW}}$ and becomes substantially enhanced only at considerably lower temperatures. To isolate the intrinsic electronic evolution associated with the primary CDW onset, we therefore restrict our analysis to temperatures above this secondary regime. Our results thus constrain excitonic insulator scenarios that require a temperature driven gap opening at $T_{\mathrm{CDW}}$, and distinguish the primary CDW transition from deeper low-temperature reconstructions.

\section*{Acknowledgments}

This research used resources at the ESM (21-ID-1) beamline of the National Synchrotron Light Source II, a U.S. Department of Energy (DOE) Office of Science User Facility operated by Brookhaven National Laboratory under Contract No.~DE-SC0012704. This research also used the Center for Functional Nanomaterials which are U.S. Department of Energy (DOE) Office of Science User Facilities operated by Brookhaven National Laboratory under Contract No. DE-SC0012704. Author T.Y. notes that this research also benefited from support provided by the Xiamen University Malaysia research grant (Grant No.~IPHY/0008). Author S.M.H also notes that the STM experiments were supported by the National Science Foundation DMR 2226098.

\bibliographystyle{apsrev4-2}
\bibliography{References}

\end{document}